\documentclass[twocolumn,prd,nofootinbib,aps,floats,floatfix,amsmath,amssymb,secnumarabic]{revtex4} %
\usepackage{graphicx}
\usepackage[usenames,dvipsnames]{color}
\usepackage[colorlinks,citecolor=blue]{hyperref}
\usepackage{amsmath}
\usepackage{bbm}
\usepackage{amsfonts}
\usepackage{amssymb}
\usepackage{latexsym}
\usepackage{graphicx}
\usepackage[english]{babel}
\usepackage{multirow}
\usepackage{float}
\usepackage{url}
\usepackage{slashed}
\usepackage{xcolor} 

\newcommand{\be}{\begin{equation}}
\newcommand{\ee}{\end{equation}}
\newcommand{\ba}{\begin{array}}
\newcommand{\ea}{\end{array}}
\newcommand{\bea}{\begin{eqnarray}}
\newcommand{\eea}{\end{eqnarray}}

\usepackage{ulem,fancyvrb}
\usepackage{xcolor}

\begin{document}

\title{TeV dark matter and the DAMPE electron excess}
\author{Xuewen Liu}\email{xuewenliu@nju.edu.cn}
\author{Zuowei Liu}\email{zuoweiliu@nju.edu.cn}
\affiliation{School of Physics, Nanjing University, Nanjing, 210093, China}

\begin{abstract}  
The recent high energy electron and positron flux observed by the DAMPE experiment 
indicates possible excess events near 1.4 TeV. Such an excess may be evidence of  
dark matter annihilations or decays in a dark matter subhalo 
that is located close to the solar system. We give here an analysis of this excess 
from annihilations of Dirac fermion dark matter which is 
charged under a new $U(1)_X$ gauge symmetry. 
The interactions between dark matter and the standard model particles  
are mediated the $U(1)_X$ gauge boson. 
We show that dark matter annihilations from a local subhalo  
can explain the excess with the canonical thermal annihilation 
cross section. 
We further discuss the constraints from the relic density, from the 
dark matter direct detection, from the dark matter indirect detection, from 
the cosmic microwave background, and from the particle colliders.

\end{abstract}
\maketitle

\section{Introduction}
\label{sec:introduction}

Recently, evidence for excess electron and positron events near 1.4 TeV has been 
reported by the DAMPE experiment 
\cite{Ambrosi:2017wek}. 
Such an excess was not found previously 
by the AMS-02  
\cite{Accardo:2014lma, Aguilar:2014mma, Aguilar:2014fea} 
and by the Fermi-LAT \cite{Abdollahi:2017nat}. 
In the recent paper by the CALET experiment \cite{Adriani:2017}, 
the electron and positron events in the two energy bins near 1 TeV 
also appear higher than expected. 
Due to the better energy resolution of the DAMPE 
experiment ($\sim 1\%$ for 1 TeV electrons and positrons) than  
AMS-02, Fermi-LAT, and CALET, the TeV electron and positron flux 
can be measured with better accuracy by the DAMPE experiment 
than measured previously by other experiments. 
In the DAMPE data, 
the electron and positron excess events occur only in one energy bin near 1.4 TeV. 
The localized feature in the energy spectrum of the excess events 
hints a nearby source of the 
high energy electrons and positrons. 
Inspired by this excess, studies with dark matter (DM) explanation 
\cite{Fan:2017sor,Duan:2017pkq,Gu:2017gle} 
and with astrophysical explanation 
\cite{Yuan:2017ysv, Fang:2017tvj}
have been carried out. 
In this paper, 
we study the possibility of attributing the excess of the 
TeV electrons and positrons to the DM annihilations in the 
vicinity of the solar system. 
%%

%% model

\section{The model}
\label{sec:model}

We consider a $U(1)_X$ extension of the standard model (SM)
with $X_\mu$ as the new gauge boson, and 
$\chi$ as the Dirac DM particles  
which is charged under the $U(1)_X$ gauge symmetry. 
The new Lagrangian terms are  
\be
{\cal L} = -{1\over 4} X_{\mu\nu} X^{\mu\nu} 
-{1\over 2} M_X^2 X_\mu X^\mu 
+X_\mu J^\mu, 
\ee
where $X_\mu$ is the new gauge boson, 
$X_{\mu\nu}$ is the field strength, 
$M_X$ is the $X$ boson mass. 
The $X$ boson is assumed to couple to the DM 
fermions and to the Dirac DM in the vector 
current form, $J_\mu = g_f \bar f \gamma_\mu f 
+ g_\chi \bar \chi \gamma^\mu \chi$.

In this paper, we consider three scenarios: 
(1) DM annihilates into $e^+e^-$ only via an s-channel 
exchange of the $X_\mu$ boson (assuming only $g_e \neq 0$); 
(2) DM annihilates into all SM final states universally via the 
s-channel $X_\mu$ boson; 
(3) DM annihilates into a pair of on-shell $X$ bosons which 
subsequently decay into SM fermions.

 The annihilation cross section for the $\chi\chi\to X \to \bar f f$ process is given 
 by 
 \be
\sigma v = {N_f g_\chi^2 g_f^2 \over 6 \pi} 
{s+2m_\chi^2 \over (s-M_X^2)^2 + M_X^2 \Gamma_X^2}
 \ee
where $N_f = 1(3)$ for leptons (quarks), 
$\Gamma_X$ is the total decay width of the $X$ boson, 
and we have neglected the mass of the final state SM particle. 
The partial decay with of the $X$ boson is 
$
\Gamma (X\to \bar f f) = N_f g_f^2 M_X/(12\pi). 
$
If the DM annihilation occurs away from the $X$ resonance 
and the $X$ boson has a narrow decay width, i.e.\ 
$\Gamma_X \ll M_X$, one has  
$
\sigma v \simeq  
{N_c^f g_f^{2} g_\chi^2  /(m_\chi^2\pi 
(x^2-4)^2)} 
$
where $x \equiv M_X/m_\chi$. 
For the case in which the $X$ boson only couples to electrons, 
in order to obtain the desired 1 pb annihilation cross section, 
one has 
$
\sqrt{|g_e g_\chi|}  \simeq 0.37 \sqrt{|x^2-4|}
$
where $x = M_X/(1.5~\text{TeV})$, 
if the DM annihilation occurs away from the $X$ resonance.

%% propagation 

\section{Cosmic Ray Propagation}
\label{sec:propagation}

The propagation of the electrons and positrons can 
be described by the following diffusion equation 
\be
\partial_ t f - \partial_E (b(E) f) - D(E) \nabla^2 f  = Q
\ee
where $f = {dN/(dE\, dV)}$ is the electron energy spectrum, 
$b(E) = -dE/dt$ is the energy loss coefficient, 
$D(E)$ is the diffusion coefficient, 
and $Q=Q({\bf x}, E, t)$ is the source term. 
We take $b(E)=b_0 (E/\text{GeV})^2$ with $b_0=10^{-16}$ GeV/s 
and $D(E)=D_0(E/\text{GeV})^\delta$. 
The diffusion coefficient $D(E)$ depends on the 
height $2L$ in the $z$ direction of the cylindrical 
diffusion zone which is usually assumed for the Milky Way galaxy. 
We adopt the medium case in Ref.\ \cite{Cirelli:2008id}  
such that  $L=4$ kpc,  $D_0$ = 11 pc$^2$/kyr, and $\delta=0.7$. 
For the steady-state case, the Green function of the diffusion 
equation is given by  
\cite{Ginzburg, Kuhlen:2009is, Delahaye:2010ji}
\be
G({\bf x}, E; {\bf x}_s, E_s) = {\exp\left[-{({\bf x-x}_s)^2/\lambda^2}\right]
\over b(E) (\pi \lambda^2)^{3/2}}, 
\ee
where $\lambda$ is the propagation scale which is given by 
\be
\lambda^2 = 4 \int_E^{E_s} dE' {D(E') \over b(E')}. 
\ee
By using the Green function, the general solution to the diffusion 
equation can be computed via 
\be
f({\bf x}, E) = \int d^3 x_s \int dE_s\, 
G({\bf x}, E; {\bf x}_s, E_s) Q({\bf x}_s, E_s). 
\ee
For DM annihilations, the source function of electrons and 
positrons is given by 
\be
Q({\bf x}, E) = {1 \over 4} { \rho^2_\chi({\bf x}) \over m_\chi^2 } 
\langle \sigma v \rangle {dN \over dE}, 
\ee
where $ \rho_\chi({\bf x})$ is the DM mass density, 
$m_\chi$ is the DM mass, 
$\langle \sigma v \rangle$ is the velocity-averaged 
annihilation cross section,  
${dN \over dE}$ is the energy spectrum of 
electrons and positrons per annihilation, 
and the $1/4$ factor is due to the Dirac nature of the DM
($1/2$ for Majorana DM). 
The electron and positron flux per unit energy from DM annihilations 
is given by $\Phi({\bf x}, E) = v f({\bf x}, E)/(4\pi)$ where 
$v$ is the electron velocity; the unit of the flux is 
1/ (GeV $\cdot$ cm$^2$ $\cdot$ s $\cdot$ sr) 
\cite{Cirelli:2008id}.

%% subhalo  

\section{electron flux from a local subhalo}
\label{sec:fitting}

We consider an ultra-faint DM subhalo which 
is located $\lesssim$1 kpc away from us. 
We assume a NFW density profile \cite{Navarro:1996gj} 
for the subhalo  
\be
\rho(r) = \rho_s {(r/r_s)^{-\gamma} \over (1+r/r_s)^{3-\gamma}}. 
\ee
The distance between the subhalo and us is denoted 
as $d_s$ which is taken to be in the range (0.1-1) kpc. 
In the following we consider two different sets of parameters 
$(\gamma, \rho_s, r_s, d_s) = (1, 1, 1,1)$ (denoted as SHA) and $(0.5, 100, 0.1,0.3)$ 
(denoted as SHB) 
where $\rho_s$ is in GeV/cm$^3$, and  
$r_s$ and $d_s$ are in kpc 
\cite{Choquette:2017nqk, Walker:2008ax, Simon:2007dq}.

If two DM particles with 1.5 TeV mass annihilate in the subhalo via $\chi\chi \to e^+e^-$ only 
with $\sigma v = 3\times 10^{-26}$ cm$^3$/s, 
the electron flux at $E=1.4$ TeV is 
$E^3 \Psi_{\chi} = 0.2 (47) $ GeV$^2$/(m$^2$ $\cdot$ s $\cdot$ sr) 
for the SHA (SHB) case. 
Thus, in order to produce the right amount of 
electrons and positrons observed by the DAMPE experiment, 
we will adopt the SHB assumption throughout the paper.

%% background

\section{Cosmic Ray Background}
\label{sec:bkg}

To understand the cosmic ray (CR) background is essential for 
astrophysical observations. 
To know the background predictions precisely in the measurement of the 
satellite experiments is the major challenge of the current DM 
indirect search experiments.    
%%%%
Usually the CR background can be modelled via the broken power-law forms. 
We adopt the parameterization formulas as in Ref.~\cite{Huang:2016pxg} 
where the background electrons and positrons consists of three components:  
the primary electrons from the CR sources, $\phi_\text{primary}$, 
the secondary positrons/electrons originating from interactions 
between the primary CR and the interstellar medium, $\phi_\text{secondary}$, 
and the extra source, for examples pulsars or DM, $\phi_\text{source}$. 
For the primary electrons, the flux is parameterized as 
$
\phi_\text{primary} = C E^{-\alpha}/(1+(E/E_b)^{\beta}) 
$; 
the secondary positron flux takes the same formula  
as the primary electron but with different coefficients. 
The extra sources contains an exponential cut-off scale $E_c$, 
which takes the form 
$
\phi_\text{source} = C E^{-\gamma}\exp(-E/E_c)
$.
The total flux for positrons is given by 
$
\Phi_{e^+} = \phi_\text{secondary} + \phi_\text{source} 
$, 
and the total flux for electrons is given by 
$
\Phi_{e^-} = \phi_\text{primary} + 0.6\, \phi_\text{secondary} + \phi_\text{source} 
$ \cite{Huang:2016pxg}. 
In our analysis, we used the electron plus positron flux measured in DAMPE to fit the 
various coefficients. 
The best fit model is given as follows: 
for the primary electrons, 
($C=16.67 $ GeV$^2$ m$^{-2}$ s$^{-1}$ sr$^{-1}$ ,  
$\alpha=1.20, 
\beta=2.10,  
E_b= 4.98$ GeV);  
for the secondary electron/positron flux, 
($C= 0.80 $ GeV$^2$ m$^{-2}$ s$^{-1}$ sr$^{-1}$,  
$\alpha=0.76, 
 \beta=2.51,  
 E_b= 1.40$ GeV);  
 for the additional source, 
($C=0.99$ GeV$^2$ m$^{-2}$ s$^{-1}$ sr$^{-1}$, 
 $\gamma=2.32,  E_c=686.75$ GeV).

%% propagation 

\section{DAMPE data fitting}
\label{sec:fitting}

We use $\Psi_{B}+\Psi_{\chi}$ to fit the DAMPE data, 
where $\Psi_{B}$ is the CR background, 
and $\Psi_{\chi}$ is the DM contribution from both the Milky Way 
(MW) halo and the nearby subhalo. 
We carry out a $\chi^2$ analysis  
\begin{eqnarray}
\chi^2=\sum_{i}\frac{(E_i^3 \Phi_i^{\text{th}} - E_i^3 \Phi_i^{\text{exp}} )^2}{\delta_i^2},\label{eq:chi2}
\end{eqnarray}
where $\Phi_i^{\text{th}}= \Psi_{B} + \Psi_{\chi}$ is the predicted spectrum
of electron plus positron, $\Phi_i^{\text{exp}}$ is the experiment data 
observed by the DAMPE experiment, and 
$\delta_i$ is the uncertainties reported by the DAMPE experiment. 
For the DM signal, we use PPPC4DM ID 
\cite{Cirelli:2010xx,Ciafaloni:2010ti} to generate the 
energy spectrum for the source function.

\begin{figure}[t]
\centerline{
\includegraphics[width=0.85\columnwidth]{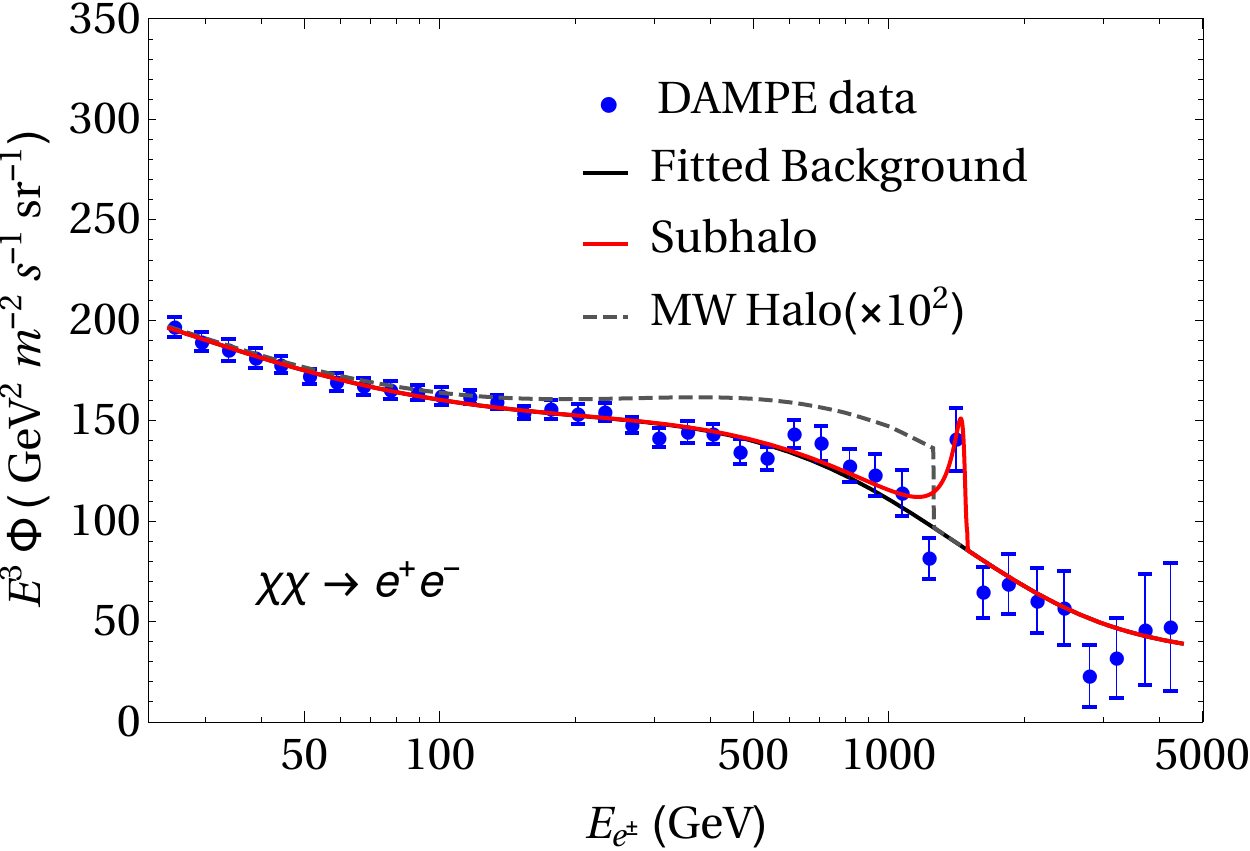}
}
\caption{The DAMPE electron data and the DM 
signals via $\chi\chi\to e^+e^-$.   
The black line represents the fitted background; 
the red line is the contribution  
from a local DM subhalo;  
The dashed gray line is contribution  
from the Milky Way halo. 
$m_\chi = $ 1.5 TeV and  
$ \langle \sigma v \rangle = 3 \times 10^{-26}$ cm$^3$/s.  
}
\label{fig1}
\end{figure}

Fig.~(\ref{fig1})  shows the DMAPE data and the various contributions to the 
electron flux. Here we analyze the    
$\chi \chi \to X \to e^-e^+$ annihilation channel only, 
and a delta function energy spectrum
$dN/dE=2\delta(E-m_\chi)$ is employed for the injection source. 
The DM annihilation cross section  
is taken to be $\sigma v = 3\times 10^{-26}$ cm$^3$/s and mass $m_\chi=1.5$ TeV. 
As shown in Fig.~(\ref{fig1}), the DAMPE excess events are well fitted by 
the hard spectrum from a local DM subhalo if DM pair-annihilates into $e^-e^+$. 
For comparison, we also calculate the electron and positron flux from DM 
annihilations in the MW halo, which, however, 
is two order of magnitude smaller and exhibits a rather flat spectrum over a 
much larger energy range.

\begin{figure}[t]
\centerline{
\includegraphics[width=0.85\columnwidth]{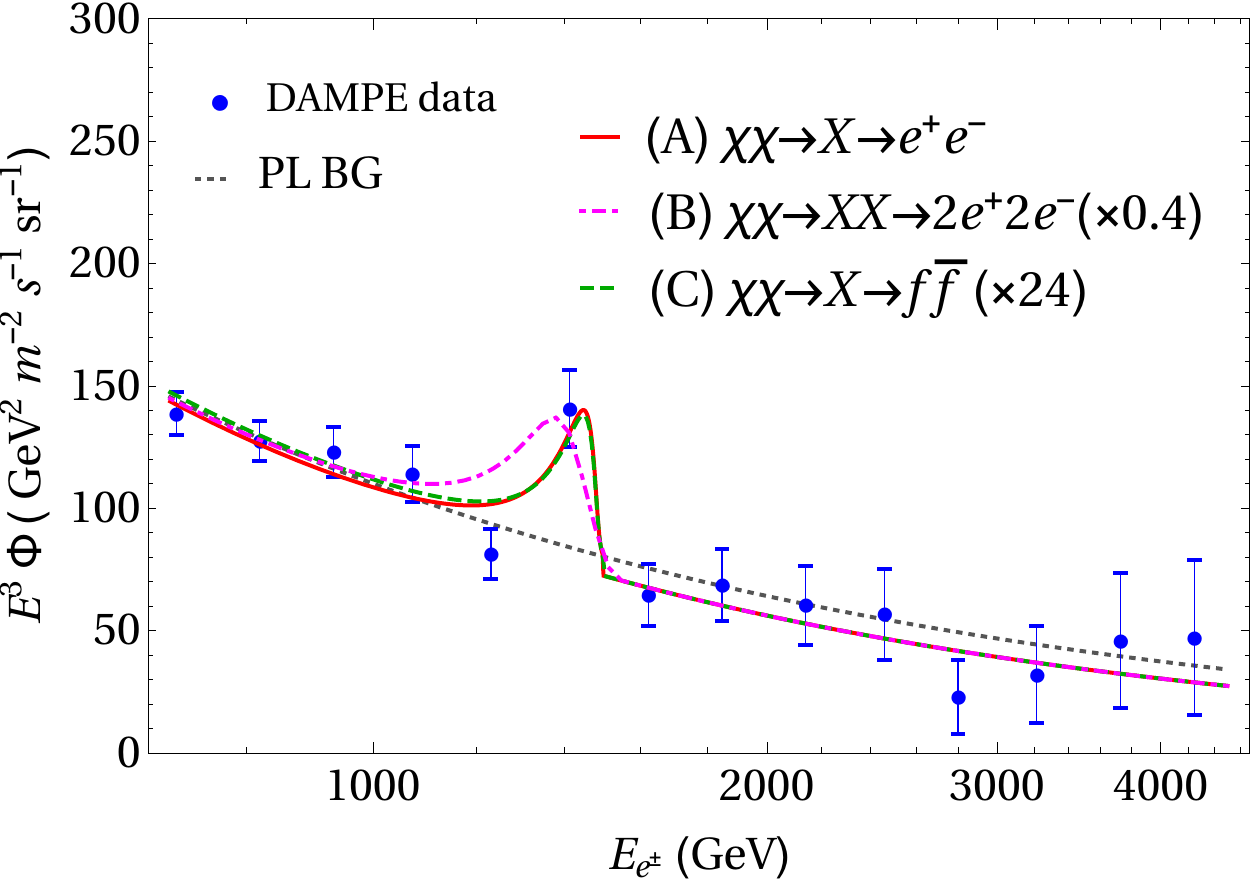}
}
\caption{The 14 high energy bins in the DAMPE data and 
contributions from various DM models. 
The background is given by a simple power law $C E^{-\gamma}$ with 
$C=2.4\, (4.7 )\times 10^4$ GeV$^2$ m$^{-2}$ s$^{-1}$ sr$^{-1}$ 
and $\gamma=0.78\, (0.89)$, 
for the background (background plus signal) fitting. 
The red solid, 
magenta dash-dotted and 
green dashed lines correspond to the DM contributions from 
the processes 
$\chi\chi\to X \to e^+ e^-$, 
$\chi\chi\to XX \to 2e^+ 2e^-$, and 
$\chi\chi\to X \to \bar f f $ respectively. 
The annihilation cross section is 
$ \langle \sigma v \rangle = 3 \times 10^{-26}$ cm$^3$/s, and 
the DM mass is taken to be 1.5 TeV for $e^+e^-$ and $\bar f f$ cases, 3 TeV for 
the $4e$ case. 
To generate the observed electron excess, an overall factor of 24 (0.4) 
is multiplied for the $\bar f f$ ($4e$) channel.
}
\label{fig2}
\end{figure}

In Fig.~(\ref{fig2}), we overlay signals from different DM models with 
the 14 high energy bins in the DAMPE data. 
Instead of using the global fitting background used in Fig.~(\ref{fig1}), 
a simple power law (PL) background $C E^{-\gamma}$ is employed for the TeV electrons. 
%to parameterize the background at the TeV energy scale.  
The best fit PL has 
$C=2.4 \times 10^4$ GeV$^2$ m$^{-2}$ s$^{-1}$ sr$^{-1}$ 
and $\gamma=0.78$ and the minimum $\chi^2=20.4$. 
We consider three different annihilation modes: 
($A$) $\chi\chi\to X\to e^+e^-$, 
($B$) $\chi\chi\to XX\to 2e^+2e^-$,  
($C$) $\chi\chi\to X\to \bar f f$. 
In the case $A$ and the case $B$, the $X$ boson couples only electrons; 
in the case $C$, the $X$ boson couples to all SM fermions with 
universal couplings.

 %%%
 %%%
For different DM annihilation channels, we employ  $\Phi_B + F \Phi^{0}_{\chi}$ to fit the 14 high energy data points, 
where 
$\Phi_B= C E^{-\gamma}$, 
$\Phi^{0}_{\chi}$ is the flux corresponding to the canonical thermal cross section $\sigma v = 3\times 10^{-26}$ cm$^3$/s, 
$F$ is a overall floating parameter. 
The DM mass is fixed at 1.5 TeV for case $A$ and $C$, 3 TeV for case $B$. 
For case $A$, we find  
$C=4.7 \times 10^4$ GeV$^2$ m$^{-2}$ s$^{-1}$ sr$^{-1}$, $\gamma=0.89$ and $F=1.03$, 
with the minimum $\chi^2=$ 9.0 which is improved by $\Delta\chi^2=11.4$ 
from the background-only PL fitting. 
%%%
%%%
In the case $B$, $\chi\chi\to XX\to 2e^+2e^-$, the energy spectrum of electron and positron is 
a box-shaped distribution 
\cite{Ibarra:2012dw, Mardon:2009rc, Abdullah:2014lla, Cline:2014dwa, Agrawal:2014oha, Cline:2015qha}
which depends on 
the mass gap $\Delta m= m_\chi- M_X$ 
which should be small to explain the sharp energy spectrum in the DAMPE data. 
We take $\Delta m= 2$ GeV and $F=0.4$ which has $\chi^2=$ 15.3 ($\Delta\chi^2=5.1$).
%%%
%%%
In the case $C$, $\chi\chi\to X\to \bar f f$, we used PPPC4DMID to compute the energy spectrum. 
In our model, the branching ratio $BR = 1/24(1/8)$ for each lepton (quark) final state. 
We find that  $\chi^2$ is 9.4 ( $\Delta\chi^2=11$) for $F=24$. 
The flux in the case $C$ overlaps with case $A$, indicating that the 
electron channel is the dominant mode.

 %% collider constraints  

\section{Collider constraints}
\label{sec:constraints}

If the $X$ boson couples to both quarks and leptons, 
one can also search for the resonance in the dilepton signal  
at the LHC. 
ATLAS collaboration also used the four-fermion 
contact interaction Lagrangian to interpret the 
most recent LHC results on dilepton signals \cite{Aaboud:2017buh}
\be
{\cal L} = {4\pi \over \Lambda^2} \eta_{ij} 
(\bar q_i \gamma^\mu q_i)
(\bar \ell_j \gamma^\mu \ell_j), 
\ee
where $i, j = L, R$ for different chiral interactions, 
and $\eta_{ij} = \pm$ denotes constructive and 
destructive interferences with the DM 
processes. 
In our model, $\Lambda = \sqrt{4\pi}M_X/\sqrt{| g_qg_\ell| }$. 
The most stringent 95\% C.L.\ lower limit on $\Lambda$ 
analyzed using the recent ATLAS data \cite{Aaboud:2017buh} 
comes from the $\ell\ell$ final state for the $LL$ chiral interaction in the 
constructive case, $\Lambda > 40$ TeV, which gives rise to 
a constraint 
$
\sqrt{|g_qg_\ell |}<0.09  (M_X/\text{TeV}). 
$

The LEPII experiment also constrain the new physics models 
via the contact interaction operators, in which the 
constraint is expressed in terms of lower bound on the 
new physics scale $\Lambda$. 
The LEPII group 
finds that $\Lambda_{VV}>15.9$ TeV in the 
$e^-e^+\to e^-e^+$ process at 95 \% C.L.\ 
\cite{LEP:2003aa} \cite{Schael:2006wu}. 
The theoretical value of  $\Lambda_{VV}$ predicted in our 
model is 
$
\Lambda_{VV}  = {\sqrt{\pi} M_X / g_e}
$; 
thus the constraints on the electron coupling is given 
$
g_e \lesssim 0.11 (M_X/\text{TeV})
$.

 %% gamma ray constraints  

\section{DM direct detection constraints}
\label{sec:dd}

PandaX experiment \cite{Cui:2017nnn} 
provides the best constraints to DM-proton 
spin-independent (SI) cross section for the 1.5 TeV DM, 
$
\sigma_{\chi p}^\text{SI} \lesssim 1.7 \times 10^{-45}~\text{cm}^2
$. 
The theoretical prediction of the SI DM-proton cross section in our model is 
$
\sigma_{\chi p} = \mu_{\chi p}^2 g_\chi ^2g_p^2/(\pi M_X^4) 
$
where 
$
g_p = 2 g_u + g_d
$. 
The above SI cross section upper limit provides a constraint on the 
coupling 
$
\sqrt{|g_\chi g_p|} \lesssim 6 \times 10^{-2} (M_X/\text{TeV}). 
$

If the $X$ boson only couples to the SM electrons, we need consider the 
DM direct detection limits due to electron recoils. 
Ref.\ \cite{Essig:2017kqs} analyzed the Xenon10 and Xenon100 
results to constrain the interaction cross section between 
DM and electron; for the 1.5 TeV DM, the upper bound on the 
cross section between DM and the free electron is 
$\sigma_{\chi e} < 3 \times 10^{-38}$ cm$^2$. 
The theoretical prediction of the DM-electron cross section in our model is 
$
\sigma_{\chi e} = \mu_{\chi e}^2 g_\chi ^2g_e^2/(\pi M_X^4) 
$
\cite{Essig:2017kqs}. 
Thus the upper bound on the couplings is 
$
\sqrt{|g_\chi g_e|} \lesssim 1.8 \times 10^{2} (M_X/\text{TeV}). 
$

To interpret the direct limit, 
we only assume the DM contribution 
from the MW halo. Although the DM density from the 
subhalo can be significant, its contribution to the DM 
direct detection signal is offset by the smaller velocity dispersion 
in the subhalo and the lack of the relative motion with respect to Earth, 
since the subhalo is also assumed to rotate around the Galactic Center.

 %% DM indirect detection constraints  

\section{DM indirect detection constraints}
\label{sec:gamma}

%% gamma ray constraints  

For the 1.5 TeV DM annihilating into two SM particles, 
H.E.S.S.\ data \cite{Abdallah:2016ygi} constrain the annihilation 
cross sections: 
$
\langle  \sigma v \rangle < 2 (6) \times 10^{-26}~\text{cm}^3/s 
$
for the $\tau^+\tau^-$ ($W^+W^-$) channel, which is based on 
10 years of the inner 300 pc of the Galactic Center region assuming 
the Einasto profile. This provides a very strong constraint 
on the case where the $X$ boson couple universally to all 
SM fermions.

For the pair-annihilation case, $\chi\chi \to XX$, 
the most stringent constraint comes from the H.E.S.S.\ data 
\cite{Profumo:2017obk} 
\cite{Abdallah:2016ygi}, 
which is stronger than the limits from the Fermi-LAT data 
in the direction of the dwarf spheroidal galaxies 
\cite{Ackermann:2015zua}
and from the Planck CMB data which is sensitive to 
energy injection to the CMB from DM annihilations 
\cite{Slatyer:2015jla} 
\cite{Slatyer:2015kla}. 
The H.E.S.S.\ limit in the $\chi \chi \to 4e$ channel is 
$
\langle  \sigma v \rangle < 6 (20) \times 10^{-25}~\text{cm}^3/s 
$
in the $m_\chi \sim M_X$ ($m_\chi \gg M_X$) case, 
for the 3 TeV DM. 
The required DM annihilation cross section for the 
DMAPE excess in the $4e$ case is  
consistent with the above H.E.S.S.\ constraint.

%%   neutrinos 

The IceCube experiment  \cite{Aartsen:2017ulx} sets an upper bound on 
the annihilation cross section for various SM channels 
by analyzing the neutrino signal 
$
\langle  \sigma v \rangle < {\cal O}(1) \times 10^{-22}~\text{cm}^3/s, 
$
which is much larger than the annihilation cross section 
needed for the DAMPE excess.

\section{Parameter space}

For the electrophilic case in which the $X$ boson 
only couples to the electrons,  
to satisfy the LEPII and direct detection constraints, 
we select a benchmark model point as follows 
$
(\delta m, g_e, g_\chi) = (100~\text{GeV}, 0.1, 0.4)
$
where $\delta m \equiv  2 m_\chi - M_X$. 
The DM annihilation for this benchmark model  is 
$
\sigma v \simeq 1.3~\text{pb}
$
which can explain the DMAPE excess and relic density.

\begin{figure}[htbp]
\centerline{
\includegraphics[width=0.85\columnwidth]{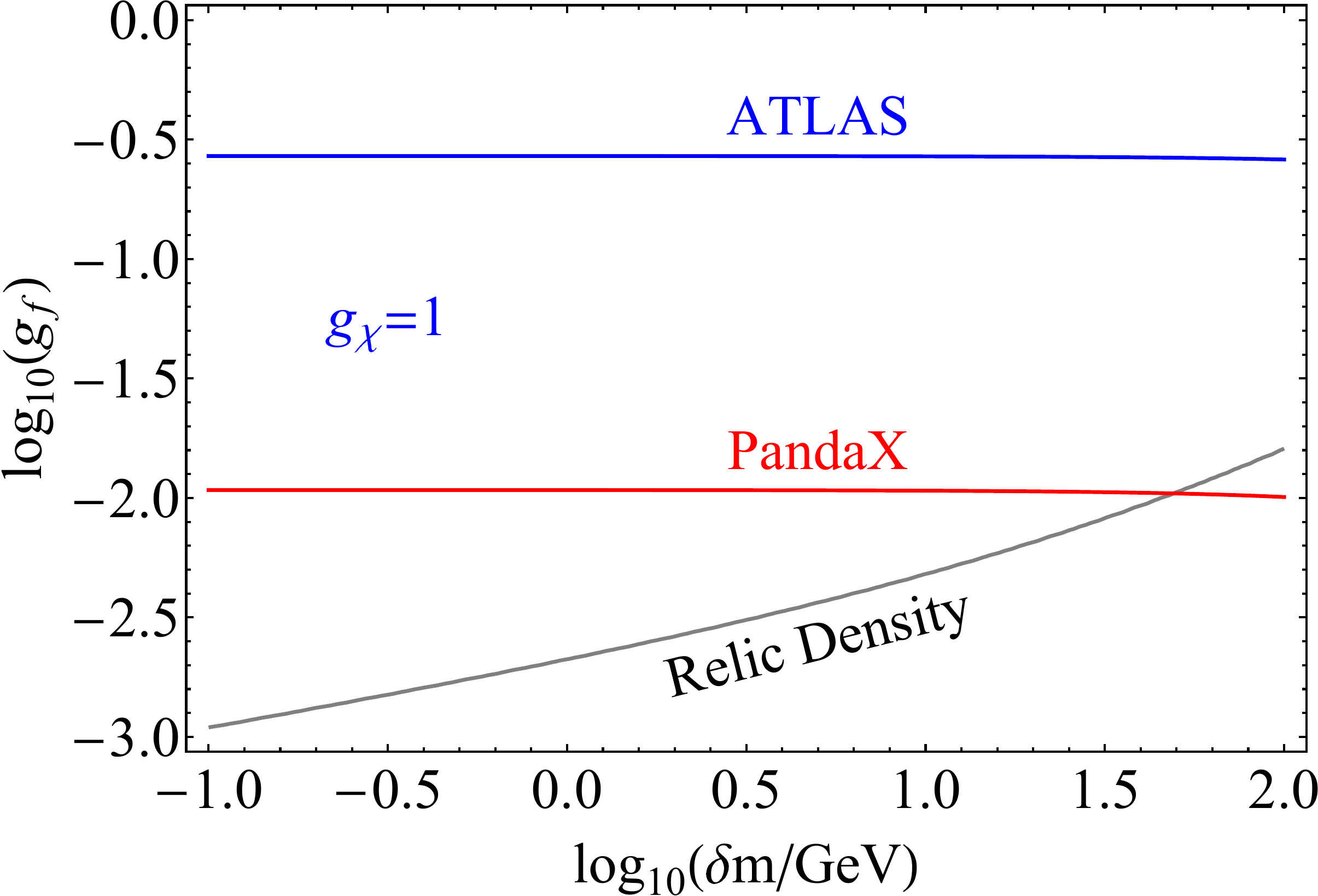}
}
\caption{Constraints on the parameter space ($\delta m,  g_f$) for the $g_\chi=1$ case. 
The blue, red lines correspond to the ATLAS and PandaX upper bounds; 
the black line indicates the parameter space that can generate the correct DM relic density.
}
\label{fig3}
\end{figure}

For the universal case in which the $X$ boson 
couples to all SM fermions with equal coupling strength, 
we select a benchmark model point as follows 
$
(\delta m, g_f, g_\chi) = (10~\text{GeV}, 4 \times 10^{-3}, 1)
$. 
The DM annihilation at the halo is 
$
\sigma v \simeq 30~\text{pb}
$; 
the DM annihilation at $v=1/4$ (the typical temperature for 
DM thermal freeze-out) is 
$
\sigma v \simeq 0.3 ~\text{pb}
$
which is smaller than 1 pb. 
However, since the annihilation cross section depends on 
the velocity of the DM, one has to take into account the thermal average 
at the freeze-out. 
Such an enhancement of the halo annihilation was 
studied previously in the context of PAMELA positron excess 
\cite{Feldman:2008xs,Ibe:2008ye,Guo:2009aj}. 
Fig.\ (\ref{fig3}) shows the relic density line and the ATLAS and PandaX constraints 
for the universal $X$ case near the resonance region. 
The benchmark model point is consistent with all constraints as shown in Fig.\ (\ref{fig3}). 
The universal case is in tension with the H.E.S.S.\ 
constraint for the $\tau^+\tau^-$ channel with an Einasto profile. 
However, for a different DM profile, the tension with the H.E.S.S.\ constraint 
could be alleviated.

For the case where DM annihilates into a pair of on-shell $X$ bosons, 
since the coupling between the $X$ boson and the SM particles can be significantly small, 
the only relevant constraint comes from the indirect detection limits. 
In our case, the most stringent  constraint comes from the H.E.S.S.\ data, 
which, however, is one order of magnitude higher than the thermal annihilation 
cross section.  
In order to produce a narrow energy spectrum 
for the injection source function in this case, 
the mass difference between the DM and the $X$ boson 
has to be very small, which compress the 
phase space of the DM annihilation so that 
$
\sigma v (\chi \chi \to XX)
$
is very small for perturbative $g_\chi$ coupling. Thus one has to be 
in the non-perturbative region of the parameter space to generate 
a large annihilation cross section and a narrow energy spectrum for this scenario.

%% Conclusions

%\smallskip
%{\noindent\bf  Conclusions.}  

%%   
\section{Conclusions}

We have proposed a simple dark matter model to explain the high energy 
electron excess events recently observed by the DAMPE experiment. 
The morphology of the energy spectrum of the excess events hints 
a local source for the high energy electrons. 
We investigated the possibility of the DM annihilations in 
a local subhalo which is 0.3 kpc away from us to generate such an excess.

Three scenarios in the model were investigated. 
The case where DM only annihilates into $e^+e^-$ 
provides a good fitting to the excess while satisfying the various constraints. 
In the case where the $X$ boson couples universally with all SM fermions, 
DM has to annihilate near the $X$ boson resonance to 
generate a much larger annihilation cross section to 
explain the excess. 
In the case where DM annihilates into on-shell $X$ bosons, 
in order to produce the sharp excess in the energy spectrum, 
the mass gap between the DM and the $X$ boson has to be 
in the GeV range.

%% Acknowledgments

{\noindent\bf Acknowledgments.}  
We thank Lei Feng, Yue-Lin Sming Tsai, 
Qiang Yuan, Cun Zhang for helpful correspondence
or discussions. The work is supported in part  
by the Nanjing University Grant 14902303 and by the 
China Postdoctoral Science Foundation 
under Grant No.\ BX201700116.

%% thebibliography


\begin{thebibliography}{10}

%\cite{Ambrosi:2017wek}
\bibitem{Ambrosi:2017wek} 
  G.~Ambrosi {\it et al.} [DAMPE Collaboration],
  %``Direct detection of a break in the teraelectronvolt cosmic-ray spectrum of electrons and positrons,''
  doi:10.1038/nature24475
  arXiv:1711.10981 [astro-ph.HE].
  %%CITATION = doi:10.1038/nature24475;%%

%%\cite{DAMPE} 
%\bibitem{DAMPE} 
%G. Ambrosi, et al., (DAMPE Collaboration), doi:10.1038/nature24475.



%\cite{Accardo:2014lma}
\bibitem{Accardo:2014lma} 
L.~Accardo {\it et al.} [AMS Collaboration],
%``High Statistics Measurement of the Positron Fraction in Primary Cosmic Rays of 0.5–500 GeV with the Alpha Magnetic Spectrometer on the International Space Station,''
Phys.\ Rev.\ Lett.\  {\bf 113}, 121101 (2014).
%doi:10.1103/PhysRevLett.113.121101
%%CITATION = doi:10.1103/PhysRevLett.113.121101;%%
%310 citations counted in INSPIRE as of 07 Nov 2017

%\cite{Aguilar:2014mma}
\bibitem{Aguilar:2014mma} 
M.~Aguilar {\it et al.} [AMS Collaboration],
%``Electron and Positron Fluxes in Primary Cosmic Rays Measured with the Alpha Magnetic Spectrometer on the International Space Station,''
Phys.\ Rev.\ Lett.\  {\bf 113}, 121102 (2014).
%doi:10.1103/PhysRevLett.113.121102
%%CITATION = doi:10.1103/PhysRevLett.113.121102;%%
%266 citations counted in INSPIRE as of 07 Nov 2017

%\cite{Aguilar:2014fea}
\bibitem{Aguilar:2014fea} 
M.~Aguilar {\it et al.} [AMS Collaboration],
%``Precision Measurement of the ($e^+ + e^−$) Flux in Primary Cosmic Rays from 0.5 GeV to 1 TeV with the Alpha Magnetic Spectrometer on the International Space Station,''
Phys.\ Rev.\ Lett.\  {\bf 113}, 221102 (2014).
%doi:10.1103/PhysRevLett.113.221102
%%CITATION = doi:10.1103/PhysRevLett.113.221102;%%
%118 citations counted in INSPIRE as of 07 Nov 2017
 

%\cite{Abdollahi:2017nat}
\bibitem{Abdollahi:2017nat} 
  S.~Abdollahi {\it et al.} [Fermi-LAT Collaboration],
  %``Cosmic-ray electron-positron spectrum from 7 GeV to 2 TeV with the Fermi Large Area Telescope,''
  Phys.\ Rev.\ D {\bf 95}, no. 8, 082007 (2017)
  %doi:10.1103/PhysRevD.95.082007
  [arXiv:1704.07195 [astro-ph.HE]].
  %%CITATION = doi:10.1103/PhysRevD.95.082007;%%
  %6 citations counted in INSPIRE as of 05 Nov 2017
  
  
%\cite{Adriani:2017}
\bibitem{Adriani:2017} 
O. Adriani  {\it et al.} (CALET Collaboration)
Phys.\ Rev.\ Lett.\ {\bf 119} , 181101 (2017).


  
%\cite{Fan:2017sor} {Yuan:2017ysv, Fang:2017tvj,Duan:2017pkq},Gu:2017gle}
\bibitem{Fan:2017sor} 
  Y.~Z.~Fan, W.~C.~Huang, M.~Spinrath, Y.~L.~S.~Tsai and Q.~Yuan,
  %``A model explaining neutrino masses and the DAMPE cosmic ray electron excess,''
  arXiv:1711.10995 [hep-ph].
  %%CITATION = ARXIV:1711.10995;%%
  
  %\cite{Gu:2017gle}
\bibitem{Gu:2017gle} 
  P.~H.~Gu and X.~G.~He,
  %``Electrophilic dark matter with dark photon: from DAMPE to direct detection,''
  arXiv:1711.11000 [hep-ph].
  %%CITATION = ARXIV:1711.11000;%%
  
  %\cite{Duan:2017pkq}
\bibitem{Duan:2017pkq} 
  G.~H.~Duan, L.~Feng, F.~Wang, L.~Wu, J.~M.~Yang and R.~Zheng,
  %``Simplified TeV leptophilic dark matter in light of DAMPE data,''
  arXiv:1711.11012 [hep-ph].
  %%CITATION = ARXIV:1711.11012;%%
  
  %\cite{Yuan:2017ysv}
\bibitem{Yuan:2017ysv} 
  Q.~Yuan {\it et al.},
  %``Interpretations of the DAMPE electron data,''
  arXiv:1711.10989 [astro-ph.HE].
  %%CITATION = ARXIV:1711.10989;%%
  
  %\cite{Fang:2017tvj}
\bibitem{Fang:2017tvj} 
  K.~Fang, X.~J.~Bi and P.~F.~Yin,
  %``Explanation of the knee-like feature in the DAMPE cosmic $e^-+e^+$ energy spectrum,''
  arXiv:1711.10996 [astro-ph.HE].
  %%CITATION = ARXIV:1711.10996;%%
  
  
  
  %\cite{Cirelli:2008id}
\bibitem{Cirelli:2008id} 
  M.~Cirelli, R.~Franceschini and A.~Strumia,
  %``Minimal Dark Matter predictions for galactic positrons, anti-protons, photons,''
  Nucl.\ Phys.\ B {\bf 800}, 204 (2008)
  doi:10.1016/j.nuclphysb.2008.03.013
  [arXiv:0802.3378 [hep-ph]].
  %%CITATION = doi:10.1016/j.nuclphysb.2008.03.013;%%
  %145 citations counted in INSPIRE as of 29 Nov 2017
  
    
%\cite{Ginzburg}
\bibitem{Ginzburg}   
V. L. Ginzburg and S. I. Syrovatskii, The Origin of
Cosmic Rays. Pergamon, Oxford, 1964.
  
  
%\cite{Kuhlen:2009is}
\bibitem{Kuhlen:2009is} 
  M.~Kuhlen and D.~Malyshev,
  %``ATIC, PAMELA, HESS, Fermi and nearby Dark Matter subhalos,''
  Phys.\ Rev.\ D {\bf 79}, 123517 (2009)
  doi:10.1103/PhysRevD.79.123517
  [arXiv:0904.3378 [hep-ph]].
  %%CITATION = doi:10.1103/PhysRevD.79.123517;%%
  %45 citations counted in INSPIRE as of 29 Nov 2017
    
%\cite{Delahaye:2010ji}
\bibitem{Delahaye:2010ji} 
  T.~Delahaye, J.~Lavalle, R.~Lineros, F.~Donato and N.~Fornengo,
  %``Galactic electrons and positrons at the Earth:new estimate of the primary and secondary fluxes,''
  Astron.\ Astrophys.\  {\bf 524}, A51 (2010)
  doi:10.1051/0004-6361/201014225
  [arXiv:1002.1910 [astro-ph.HE]].
  %%CITATION = doi:10.1051/0004-6361/201014225;%%
  %121 citations counted in INSPIRE as of 29 Nov 2017

  
  %\cite{Navarro:1996gj}
\bibitem{Navarro:1996gj} 
J.~F.~Navarro, C.~S.~Frenk and S.~D.~M.~White,
%``A Universal density profile from hierarchical clustering,''
Astrophys.\ J.\  {\bf 490}, 493 (1997)
%doi:10.1086/304888
[astro-ph/9611107].
%%CITATION = doi:10.1086/304888;%%
%5146 citations counted in INSPIRE as of 07 Nov 2017
  
  
  %\cite{Choquette:2017nqk}
\bibitem{Choquette:2017nqk} 
  J.~Choquette,
  %``Constraining Dwarf Spheroidal Dark Matter Halos With The Galactic Center Excess,''
  arXiv:1705.09676 [astro-ph.CO].
  %%CITATION = ARXIV:1705.09676;%%

%\cite{Walker:2008ax}
\bibitem{Walker:2008ax} 
  M.~G.~Walker, M.~Mateo and E.~Olszewski,
  %``Stellar Velocities in the Carina, Fornax, Sculptor and Sextans dSph Galaxies: Data from the Magellan/MMFS Survey,''
  Astron.\ J.\  {\bf 137}, 3100 (2009)
  doi:10.1088/0004-6256/137/2/3100
  [arXiv:0811.0118 [astro-ph]].
  %%CITATION = doi:10.1088/0004-6256/137/2/3100;%%
  %80 citations counted in INSPIRE as of 29 Nov 2017
  
%\cite{Simon:2007dq}
\bibitem{Simon:2007dq} 
  J.~D.~Simon and M.~Geha,
  %``The Kinematics of the Ultra-Faint Milky Way Satellites: Solving the Missing Satellite Problem,''
  Astrophys.\ J.\  {\bf 670}, 313 (2007)
  doi:10.1086/521816
  [arXiv:0706.0516 [astro-ph]].
  %%CITATION = doi:10.1086/521816;%%
  %493 citations counted in INSPIRE as of 29 Nov 2017

  
  
 %\cite{Huang:2016pxg}
\bibitem{Huang:2016pxg} 
  X.~Huang, Y.~L.~S.~Tsai and Q.~Yuan,
  %``LikeDM: likelihood calculator of dark matter detection,''
  Comput.\ Phys.\ Commun.\  {\bf 213}, 252 (2017)
  %doi:10.1016/j.cpc.2016.12.015
  [arXiv:1603.07119 [hep-ph]].
  %%CITATION = doi:10.1016/j.cpc.2016.12.015;%%
  %4 citations counted in INSPIRE as of 05 Nov 2017

%\cite{Cirelli:2010xx}\cite{Ciafaloni:2010ti}
\bibitem{Cirelli:2010xx} 
M.~Cirelli {\it et al.},
%``PPPC 4 DM ID: A Poor Particle Physicist Cookbook for Dark Matter Indirect Detection,''
JCAP {\bf 1103}, 051 (2011)
Erratum: [JCAP {\bf 1210}, E01 (2012)]
%doi:10.1088/1475-7516/2012/10/E01, 10.1088/1475-7516/2011/03/051
[arXiv:1012.4515 [hep-ph]].
%%CITATION = doi:10.1088/1475-7516/2012/10/E01, 10.1088/1475-7516/2011/03/051;%%
%458 citations counted in INSPIRE as of 07 Nov 2017

%\cite{Ciafaloni:2010ti}
\bibitem{Ciafaloni:2010ti} 
P.~Ciafaloni, D.~Comelli, A.~Riotto, F.~Sala, A.~Strumia and A.~Urbano,
%``Weak Corrections are Relevant for Dark Matter Indirect Detection,''
JCAP {\bf 1103}, 019 (2011)
%doi:10.1088/1475-7516/2011/03/019
[arXiv:1009.0224 [hep-ph]].
%%CITATION = doi:10.1088/1475-7516/2011/03/019;%%
%201 citations counted in INSPIRE as of 07 Nov 2017  

%\cite{Ibarra:2012dw}
\bibitem{Ibarra:2012dw} 
  A.~Ibarra, S.~Lopez Gehler and M.~Pato,
  %``Dark matter constraints from box-shaped gamma-ray features,''
  JCAP {\bf 1207}, 043 (2012)
  doi:10.1088/1475-7516/2012/07/043
  [arXiv:1205.0007 [hep-ph]].
  %%CITATION = doi:10.1088/1475-7516/2012/07/043;%%
  %100 citations counted in INSPIRE as of 30 Nov 2017

%\cite{Mardon:2009rc}
\bibitem{Mardon:2009rc} 
  J.~Mardon, Y.~Nomura, D.~Stolarski and J.~Thaler,
  %``Dark Matter Signals from Cascade Annihilations,''
  JCAP {\bf 0905}, 016 (2009)
  doi:10.1088/1475-7516/2009/05/016
  [arXiv:0901.2926 [hep-ph]].
  %%CITATION = doi:10.1088/1475-7516/2009/05/016;%%
  %116 citations counted in INSPIRE as of 30 Nov 2017

%\cite{Abdullah:2014lla}
\bibitem{Abdullah:2014lla} 
  M.~Abdullah, A.~DiFranzo, A.~Rajaraman, T.~M.~P.~Tait, P.~Tanedo and A.~M.~Wijangco,
  %``Hidden on-shell mediators for the Galactic Center $\gamma$-ray excess,''
  Phys.\ Rev.\ D {\bf 90}, 035004 (2014)
  doi:10.1103/PhysRevD.90.035004
  [arXiv:1404.6528 [hep-ph]].
  %%CITATION = doi:10.1103/PhysRevD.90.035004;%%
  %130 citations counted in INSPIRE as of 30 Nov 2017
  
%\cite{Cline:2014dwa}
\bibitem{Cline:2014dwa} 
 J.~M.~Cline, G.~Dupuis, Z.~Liu and W.~Xue,
 %``The windows for kinetically mixed Z'-mediated dark matter and the galactic center gamma ray excess,''
 JHEP {\bf 1408}, 131 (2014)
 doi:10.1007/JHEP08(2014)131
 [arXiv:1405.7691 [hep-ph]].
 %%CITATION = doi:10.1007/JHEP08(2014)131;%%
 %79 citations counted in INSPIRE as of 30 Nov 2017

 %\cite{Agrawal:2014oha}
\bibitem{Agrawal:2014oha} 
 P.~Agrawal, B.~Batell, P.~J.~Fox and R.~Harnik,
 %``WIMPs at the Galactic Center,''
 JCAP {\bf 1505}, 011 (2015)
 doi:10.1088/1475-7516/2015/05/011
 [arXiv:1411.2592 [hep-ph]].
 %%CITATION = doi:10.1088/1475-7516/2015/05/011;%%
 %101 citations counted in INSPIRE as of 30 Nov 2017

%\cite{Cline:2015qha}
\bibitem{Cline:2015qha} 
 J.~M.~Cline, G.~Dupuis, Z.~Liu and W.~Xue,
 %``Multimediator models for the galactic center gamma ray excess,''
 Phys.\ Rev.\ D {\bf 91}, no. 11, 115010 (2015)
 doi:10.1103/PhysRevD.91.115010
 [arXiv:1503.08213 [hep-ph]].
 %%CITATION = doi:10.1103/PhysRevD.91.115010;%%
 %38 citations counted in INSPIRE as of 30 Nov 2017


 %\cite{Aaboud:2017buh}
\bibitem{Aaboud:2017buh} 
  M.~Aaboud {\it et al.} [ATLAS Collaboration],
  %``Search for new high-mass phenomena in the dilepton final state using 36 fb$^{−1}$ of proton-proton collision data at $ \sqrt{s}=13 $ TeV with the ATLAS detector,''
  JHEP {\bf 1710}, 182 (2017)
  doi:10.1007/JHEP10(2017)182
  [arXiv:1707.02424 [hep-ex]].
  %%CITATION = doi:10.1007/JHEP10(2017)182;%%
  %31 citations counted in INSPIRE as of 30 Nov 2017


%\cite{LEP:2003aa}
\bibitem{LEP:2003aa} 
  t.~S.~Electroweak [LEP and ALEPH and DELPHI and L3 and OPAL Collaborations and LEP Electroweak Working Group and SLD Electroweak Group and SLD Heavy Flavor Group],
  %``A Combination of preliminary electroweak measurements and constraints on the standard model,''
  hep-ex/0312023.
  %%CITATION = HEP-EX/0312023;%%
  %283 citations counted in INSPIRE as of 28 Nov 2017
  

%\cite{Schael:2006wu}
\bibitem{Schael:2006wu} 
  S.~Schael {\it et al.} [ALEPH Collaboration],
  %``Fermion pair production in $e^{+} e^{-}$ collisions at 189-209-GeV and constraints on physics beyond the standard model,''
  Eur.\ Phys.\ J.\ C {\bf 49}, 411 (2007)
  doi:10.1140/epjc/s10052-006-0156-8
  [hep-ex/0609051].
  %%CITATION = doi:10.1140/epjc/s10052-006-0156-8;%%
  %61 citations counted in INSPIRE as of 28 Nov 2017
  
  
 
%\cite{Cui:2017nnn}
\bibitem{Cui:2017nnn} 
  X.~Cui {\it et al.} [PandaX-II Collaboration],
  %``Dark Matter Results From 54-Ton-Day Exposure of PandaX-II Experiment,''
  Phys.\ Rev.\ Lett.\  {\bf 119}, no. 18, 181302 (2017)
  doi:10.1103/PhysRevLett.119.181302
  [arXiv:1708.06917 [astro-ph.CO]].
  %%CITATION = doi:10.1103/PhysRevLett.119.181302;%%
  %49 citations counted in INSPIRE as of 30 Nov 2017 
  

%\cite{Essig:2017kqs}
\bibitem{Essig:2017kqs} 
  R.~Essig, T.~Volansky and T.~T.~Yu,
  %``New Constraints and Prospects for sub-GeV Dark Matter Scattering off Electrons in Xenon,''
  Phys.\ Rev.\ D {\bf 96}, no. 4, 043017 (2017)
  doi:10.1103/PhysRevD.96.043017
  [arXiv:1703.00910 [hep-ph]].
  %%CITATION = doi:10.1103/PhysRevD.96.043017;%%
  %26 citations counted in INSPIRE as of 30 Nov 2017


%\cite{Abdallah:2016ygi}
\bibitem{Abdallah:2016ygi} 
  H.~Abdallah {\it et al.} [H.E.S.S. Collaboration],
  %``Search for dark matter annihilations towards the inner Galactic halo from 10 years of observations with H.E.S.S,''
  Phys.\ Rev.\ Lett.\  {\bf 117}, no. 11, 111301 (2016)
  doi:10.1103/PhysRevLett.117.111301
  [arXiv:1607.08142 [astro-ph.HE]].
  %%CITATION = doi:10.1103/PhysRevLett.117.111301;%%
  %55 citations counted in INSPIRE as of 29 Nov 2017


%\cite{Profumo:2017obk}
\bibitem{Profumo:2017obk} 
  S.~Profumo, F.~S.~Queiroz, J.~Silk and C.~Siqueira,
  %``Searching for Secluded Dark Matter with H.E.S.S., Fermi-LAT, and Planck,''
  arXiv:1711.03133 [hep-ph].
  %%CITATION = ARXIV:1711.03133;%%
  
  
  %\cite{Ackermann:2015zua}
\bibitem{Ackermann:2015zua} 
  M.~Ackermann {\it et al.} [Fermi-LAT Collaboration],
  %``Searching for Dark Matter Annihilation from Milky Way Dwarf Spheroidal Galaxies with Six Years of Fermi Large Area Telescope Data,''
  Phys.\ Rev.\ Lett.\  {\bf 115}, no. 23, 231301 (2015)
  doi:10.1103/PhysRevLett.115.231301
  [arXiv:1503.02641 [astro-ph.HE]].
  %%CITATION = doi:10.1103/PhysRevLett.115.231301;%%
  %499 citations counted in INSPIRE as of 29 Nov 2017

%\cite{Slatyer:2015jla}
\bibitem{Slatyer:2015jla} 
  T.~R.~Slatyer,
  %``Indirect dark matter signatures in the cosmic dark ages. I. Generalizing the bound on s-wave dark matter annihilation from Planck results,''
  Phys.\ Rev.\ D {\bf 93}, no. 2, 023527 (2016)
  doi:10.1103/PhysRevD.93.023527
  [arXiv:1506.03811 [hep-ph]].
  %%CITATION = doi:10.1103/PhysRevD.93.023527;%%
  %100 citations counted in INSPIRE as of 29 Nov 2017
  
%\cite{Slatyer:2015kla}
\bibitem{Slatyer:2015kla} 
  T.~R.~Slatyer,
  %``Indirect Dark Matter Signatures in the Cosmic Dark Ages II. Ionization, Heating and Photon Production from Arbitrary Energy Injections,''
  Phys.\ Rev.\ D {\bf 93}, no. 2, 023521 (2016)
  doi:10.1103/PhysRevD.93.023521
  [arXiv:1506.03812 [astro-ph.CO]].
  %%CITATION = doi:10.1103/PhysRevD.93.023521;%%
  %30 citations counted in INSPIRE as of 29 Nov 2017
  
%\cite{Aartsen:2017ulx}
\bibitem{Aartsen:2017ulx} 
  M.~G.~Aartsen {\it et al.} [IceCube Collaboration],
  %``Search for Neutrinos from Dark Matter Self-Annihilations in the center of the Milky Way with 3 years of IceCube/DeepCore,''
  Eur.\ Phys.\ J.\ C {\bf 77}, no. 9, 627 (2017)
  doi:10.1140/epjc/s10052-017-5213-y
  [arXiv:1705.08103 [hep-ex]].
  %%CITATION = doi:10.1140/epjc/s10052-017-5213-y;%%
  %3 citations counted in INSPIRE as of 30 Nov 2017  
   
     

%\cite{Feldman:2008xs}{Ibe:2008ye}{Guo:2009aj}
\bibitem{Feldman:2008xs} 
  D.~Feldman, Z.~Liu and P.~Nath,
  %``PAMELA Positron Excess as a Signal from the Hidden Sector,''
  Phys.\ Rev.\ D {\bf 79}, 063509 (2009)
  doi:10.1103/PhysRevD.79.063509
  [arXiv:0810.5762 [hep-ph]].
  %%CITATION = doi:10.1103/PhysRevD.79.063509;%%
  %191 citations counted in INSPIRE as of 30 Nov 2017
 
%\cite{Ibe:2008ye}{Guo:2009aj}
\bibitem{Ibe:2008ye} 
  M.~Ibe, H.~Murayama and T.~T.~Yanagida,
  %``Breit-Wigner Enhancement of Dark Matter Annihilation,''
  Phys.\ Rev.\ D {\bf 79}, 095009 (2009)
  doi:10.1103/PhysRevD.79.095009
  [arXiv:0812.0072 [hep-ph]].
  %%CITATION = doi:10.1103/PhysRevD.79.095009;%%
  %186 citations counted in INSPIRE as of 30 Nov 2017
  
%\cite{Guo:2009aj}
\bibitem{Guo:2009aj} 
  W.~L.~Guo and Y.~L.~Wu,
  %``Enhancement of Dark Matter Annihilation via Breit-Wigner Resonance,''
  Phys.\ Rev.\ D {\bf 79}, 055012 (2009)
  doi:10.1103/PhysRevD.79.055012
  [arXiv:0901.1450 [hep-ph]].
  %%CITATION = doi:10.1103/PhysRevD.79.055012;%%
  %94 citations counted in INSPIRE as of 30 Nov 2017
  
\end{thebibliography}
\end{document}